\documentclass[journal]{IEEEtran}

\usepackage{xcolor}
\usepackage{graphicx}
\usepackage{hyperref} 
\usepackage{orcidlink}

\usepackage{lipsum}
\makeatletter  
\def\@IEEEBIOphotowidth{1in}    
\def\@IEEEBIOphotodepth{0.8in}   
\def\@IEEEBIOhangwidth{1in}    
\def\@IEEEBIOhangdepth{1in}    
\makeatother

\begin{document}
%

\title{EOSC CZ: Towards the development of Czech national ecosystem for FAIR research data}

%

\author{Matej Antol \orcidlink{0000-0002-1380-5647},
        Jiří Marek \orcidlink{0000-0003-2132-762X},
        Michaela Capandová \orcidlink{0000-0001-7606-4084},
        Jaroslav Juráček \orcidlink{0000-0001-9945-2530},
        and~Luděk Matyska \orcidlink{0000-0001-6399-5453}

\thanks{Authors are with the CERIT-SC centre, Institute of Computer Science, Masaryk University, 60200 Brno, Czechia. Contact at info@eosc.cz}
\thanks{Manuscript published February 20, 2024}}

\maketitle

\begin{abstract}
This short paper presents a compact overview of the Czech approach to implementing the European Open Science Cloud and plans for developing a Czech national infrastructure for FAIR research data. Its purpose is to provide an all-encompassing summary of the near future of research data management in Czechia. As such, we deliberately attempt to explain complicated concepts in minimum words, sacrificing the precision of expression for compactness.
\end{abstract}


%
\IEEEpeerreviewmaketitle


\section{Introduction}

\IEEEPARstart{T}{he} importance of data in research is continuously rising, while approaches to store, manage and share these data seem to fall behind. The value of the data is reduced by their considerable heterogeneity and lack of structure, which leads to low reproducibility and hinders scientific progress. Open Science (OS)~\cite{open_science} seeks to address some of these current issues, focusing on data availability and sharing, urging for more collaboration and emphasising research integrity. European Open Science Cloud (EOSC)~\cite{EOSCportal, EOSC_EU} is an international initiative that builds on the Open Science principles. EOSC seeks to create a common European research environment~\cite{SRIA_MAR} to store, share and re-use research data and other digital objects without barriers. We call such data and objects FAIR~\cite{FAIR} (Findable, Accessible, Interoperable, Reusable). 


\section{EOSC CZ -- Infrastructure and Services for FAIR research data}

The establishment of fundamental principles for the Czech national EOSC implementation took place in 2021, resulting in the document called \textit{Architecture of EOSC implementation in the Czech Republic}~\cite{archiEOSC_CR}. The document represents the official start of the EOSC CZ initiative~\cite{EOSC_CZ}. The primary tangible outcome of this initiative will be a National Repository Platform (NRP) -- a core component of the National Data Infrastructure (NDI). NRP will be a federated ecosystem of distinct technological layers (see Fig.~\ref{fig:NDI}) and associated services (see below).

\begin{figure}[b]
    \centering
    \includegraphics[width=1\linewidth]{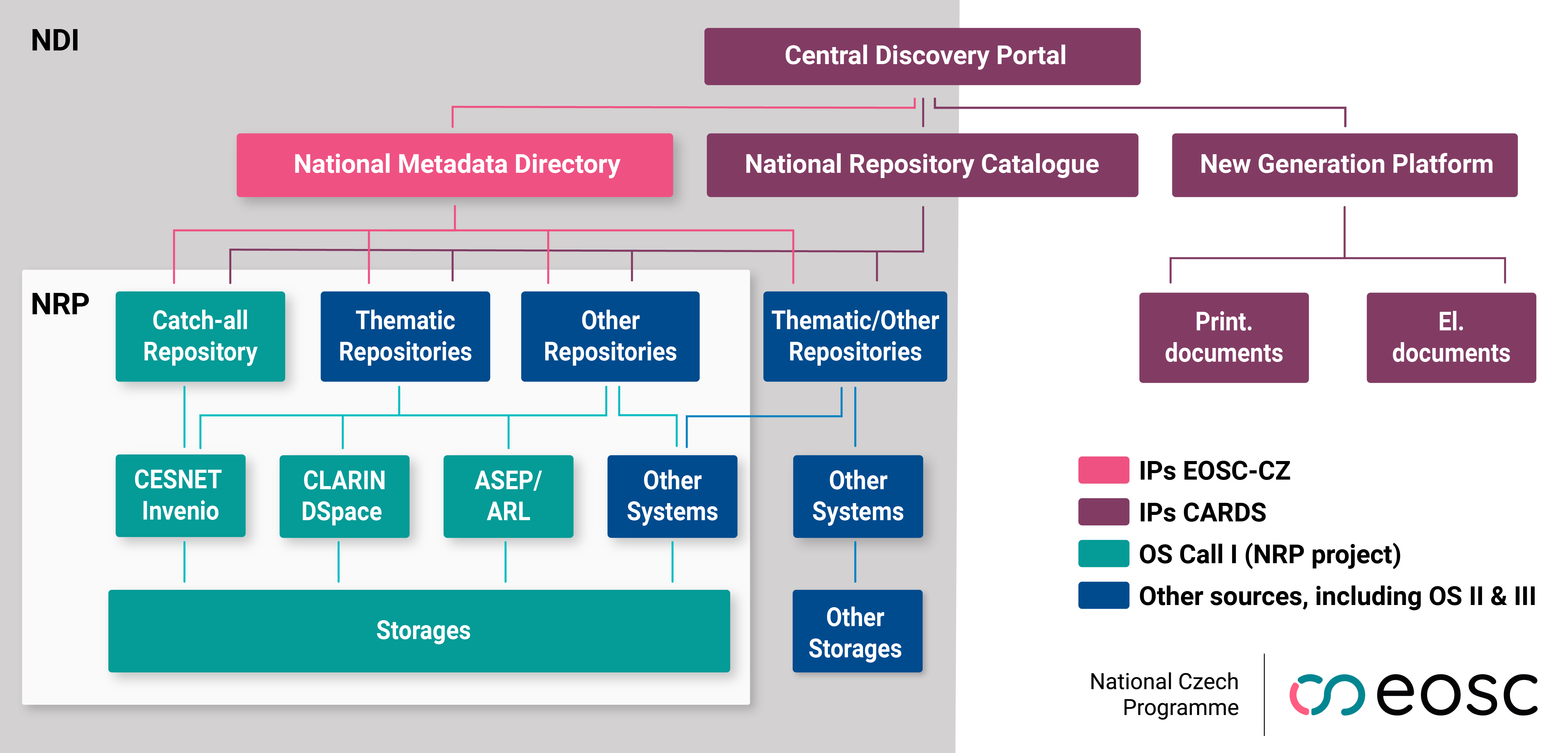}
    \caption{NDI and NRP blueprint with five abstraction layers. Bottom-up: hardware infrastructure dislocated across Czechia; three initial repository systems~-- CESNET Invenio~\cite{invenio}, CLARIN-DSpace~\cite{dspace} and ASEP/ARL~\cite{ARL}; specific domain and other repositories, metadata directory and on top, Central Discovery Portal.}
    \label{fig:NDI}
\end{figure}

The data infrastructure will complement the existing Czech national e-infrastructure e-INFRA CZ~\cite{e-INFRA} with all its services. NDI will be fully integrated at the European level~\cite{tripartite}. NRP will interconnect with the already running parts of NDI: data repositories and services held at universities, Czech Academy of Sciences and Research Infrastructures. Examples are environments such as LINDAT/CLARIAH-CZ~\cite{lindat} for natural language processing, Czech-BioImaging~\cite{Czech-BioImaging} for biological and medical imaging or EIRENE RI~\cite{eirene} for human exposome.

Next to the repositories themselves, the initiative plans to deploy and integrate several FAIR data-related services designed for NDI users. Notably:
\begin{itemize}
    \item Central Discovery Portal (CDP) integrated into the New Generation Platform (PNG) will ensure the searchability and availability of all types of resources (electronic, digitized and printed) and research results.
    \item National Metadata Directory to search in NDI metadata.
    \item Single Authentication and Authorization Infrastructure (AAI) solution Perun~\cite{perun} to guarantee data accessibility.
    \item Support for data management planning via Data Stewardship Wizard~\cite{DSW}.
    \item Support for Persistent Identifiers (PIDs)~\cite{PIDcentre}.
    \item Support for data FAIRification.
    \item Data mgmt. tools such as OneData~\cite{onedata} or iRODS~\cite{irods}.
    \item Training~\cite{training_centre} and university courses on data management.
\end{itemize}

\section{Active communities and how to participate}

Researchers' engagement is vital for the EOSC CZ's success. Since 2021, as a reaction to the EOSC CZ Architecture document, 12 EOSC CZ working groups~\cite{WGs} have been established through a self-organizing community effort. These groups will be operational during the entire EOSC CZ initiative, and registration is continuously open to new potential members. A list of their members is publicly available. Currently, the initiative is in its initial implementation phase, and the active participation of scientists in the working groups is the main guarantor for the NDI ecosystem to encompass and support all relevant research data management needs of research communities.

The initiative is also closely connected with the National Open Science coordination team within the National Library of Technology. On top of that, collaboration is being established with the already existing national Open Science communities:
\begin{itemize}
    \item Open Science working groups of the Association of Libraries of Czech Universities,
    \item national Data Steward Community and
    \item members of the institutional Open Science centres within Czech academic institutions.
\end{itemize}

\section{How to benefit from the EOSC CZ outcomes}

The NDI's ecosystem of services will be offered to the whole research community regardless of their active participation in the EOSC CZ initiative. The EOSC CZ Secretariat~\cite{secretariat} and Training Centre~\cite{training_centre} are already operational, providing consultancy, seminars and workshops for the Czech research ecosystem. The National Metadata Directory will be deployed in 2024, followed by the NRP with a portion of core services in 2025. By this time, the first domain and other repositories should also be emerging. This first phase will be completed in 2026, with an entire NRP and its services available. The initiative will concurrently foster the development of data management and other related skills for all Czech academia members. It will also encourage the systemic formation of data steward and curator roles across the academic ecosystem.

With this infrastructure, any reasonably interested Czech scientist should have sufficient information, know-how, skills, institutional support, and services to store, share, and reuse research data efficiently. 
These ambitions summarize the main objective of the EOSC CZ initiative.

\section*{Acknowledgments}
The EOSC CZ initiative has active collaborators who significantly exceed the authors of this paper. Out of these, we would namely like to acknowledge the contributions of Radka Římanová, Klára Slanařová, Petra Černohlávková, Martin Svoboda, Miroslav Bartošek, David Antoš and Michal Růžička.

\section*{Appendix: Financial support for EOSC in Czechia}

Czech Ministry of Education, Youth and Sports (MEYS) supports the EOSC CZ initiative~\cite{OP_JAK} via two systemic projects and three open science calls: 

\begin{itemize}
\item Individual Systemic Project (IPs) EOSC-CZ, coordinated by Masaryk University with two additional partners, supported with 18 mil. EUR to provide a fundamental organizational, technical, and training environment. 
\item IPs CARDS, coordinated by National Library of Technology, supported with 56 mil. EUR, to provide support for PIDs, research data description, and deliver the PNG. 
\item OS Call I, with an allocation of 50 mils. EUR, to create the NRP, its core services and related training.
\item OS Call II, with an allocation of 36 mil. EUR to support domain-specific data management, repositories and related services over the NRP. 
\item OS Call III, scope of which is currently under discussion.
\end{itemize}

\vskip -1.5\baselineskip plus -1fil


\begin{IEEEbiography}
[{\includegraphics[width=.8in,height=1in,clip,keepaspectratio]{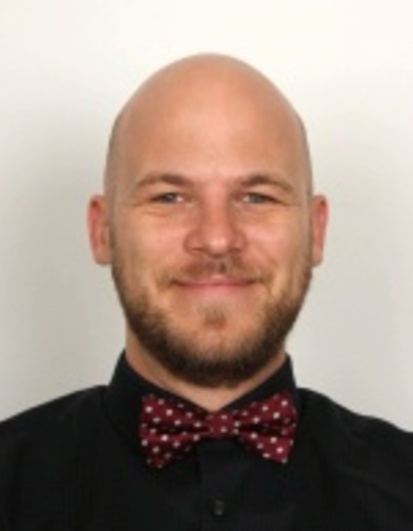}}]
{Matej Antol} is the principal project manager of the IPs EOSC-CZ. He is also the integration manager of the Czech e-infrastructure e-INFRA CZ and an executive director of one of its three partners, the CERIT-SC infrastructure. He has a long background in IT and research projects. His research activities focus on managing and analysing complex, high-dimensional data.
\end{IEEEbiography}

\vskip -2.5\baselineskip plus -1fil

\begin{IEEEbiography}
[{\includegraphics[width=.8in,height=1in,clip,keepaspectratio]{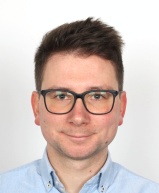}}]
{Jiri Marek} is the General Secretary of the EOSC CZ initiative and head of the EOSC CZ Secretariat. He holds the role of the Open Science manager at Masaryk University and serves as a head of the CZARMA Open Science Task Force. He is also involved with activities regarding digitization of the public sector via open technologies (Open Cities, etc.)
\end{IEEEbiography}

\vskip -2.5\baselineskip plus -1fil

\begin{IEEEbiography}
[{\includegraphics[width=.8in,height=1in,clip,keepaspectratio]{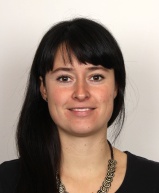}}]
{Michalea Capandova} is the secretary to the EOSC CZ Working Groups Metadata and Materials Sciences and Engineering. Her research in the biomedical field is focused on the development of cellular elements and biomaterials for lung tissue engineering. She loves electrospinning and scanning electron microscopy.
\end{IEEEbiography}


\vskip -2.5\baselineskip plus -1fil

\begin{IEEEbiography}
[{\includegraphics[width=.8in,height=1in,clip,keepaspectratio]{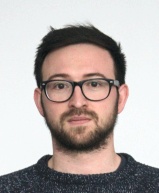}}]
{Jaroslav Juracek} is the secretary to the EOSC CZ Working Group Bio/Health/Food. Beyond that, he takes part in building the European Genomic Data Infrastructure and related activities at the national level. His focus is set on advancing open science initiatives and access to and utilization of genomic data for research and innovation.
\end{IEEEbiography}

\vskip -2.5\baselineskip plus -1fil

\begin{IEEEbiography}
[{\includegraphics[width=.8in,height=1in,clip,keepaspectratio]{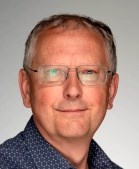}}]
{Ludek Matyska} is a full professor at the Faculty of Informatics, Masaryk University, with a long track in developing national and European research infrastructures. He is the director of the CERIT-SC, one of three members of the e-INFRA CZ steering board, the principal project manager of the NRP project, and chairman of the IPs EOSC-CZ steering committee.
\end{IEEEbiography}





\begin{thebibliography}{99}

\bibitem{open_science}
Munafò, M., Nosek, B., Bishop, D. et al. A manifesto for reproducible science. Nat Hum Behav 1 (2017). doi.org/10.1038/s41562-016-0021

\bibitem{FAIR}
Wilkinson, M.D. et al. The FAIR Guiding Principles for scientific data management and stewardship. Scientific data, 3(1), pp.1-9. (2016)

\bibitem{EOSCportal}
\url{https://eosc-portal.eu/}

\bibitem{EOSC_EU}
\url{https://eosc.eu/}

\bibitem{SRIA_MAR}
\url{https://eosc.eu/sria-mar/}

\bibitem{archiEOSC_CR}
\url{https://www.msmt.cz/uploads/311/Architektura_implementace_EOSC_v_CR.pdf}

\bibitem{EOSC_CZ}
\url{https://www.eosc.cz/en}

\bibitem{invenio}
\url{https://github.com/CESNET}

\bibitem{dspace}
\url{https://github.com/ufal/clarin-dspace}

\bibitem{ARL}
\url{https://asep-portal.lib.cas.cz/basic-information/dataset-repository/}

\bibitem{e-INFRA}
\url{https://www.e-infra.cz/en}

\bibitem{tripartite}
\url{https://eosc.eu/tripartite-collaboration/czech-republic/}

\bibitem{lindat}
\url{https://lindat.cz/}

\bibitem{Czech-BioImaging}
\url{https://www.czech-bioimaging.cz/}

\bibitem{eirene}
\url{https://www.eirene-ri.eu/}

\bibitem{perun}
\url{https://perun-aai.org/}

\bibitem{DSW}
\url{https://ds-wizard.org/}

\bibitem{PIDcentre}
\url{https://identifikatory.cz/en/}

\bibitem{onedata}
\url{https://www.cerit-sc.cz/management-of-data-workflows}

\bibitem{irods}
\url{https://irods.org/}

\bibitem{training_centre}
\url{https://www.eosc.cz/en/training-centre}



\bibitem{WGs}
\url{https://www.eosc.cz/en/working-groups}

\bibitem{secretariat}
\url{https://www.eosc.cz/en/secretariat}

\bibitem{OP_JAK}
\url{https://www.dotaceeu.cz/en/statistiky-a-analyzy/seznam-operaci-(prijemcu)}

\end{thebibliography}
\end{document}